\documentclass[11pt]{article}
\usepackage{amssymb}
\usepackage{amsmath}
\newtheorem{definition}{Definition}[section]

\newtheorem{theorem}[definition]{Theorem}

\newtheorem{remark}[definition]{Remark}

\newcommand\map[3]{#1\ \colon\ #2\longrightarrow#3}

\begin{document}
\title{A Generalization of Chetaev's Principle for a Class of Higher Order
 Non-holonomic Constraints}

\author{
Hern\'an Cendra\\
Departamento de Matem\'atica\\
Universidad Nacional del Sur, Av. Alem 1253\\
8000 Bah\'{\i}a Blanca and CONICET, Argentina.\\
{\footnotesize uscendra@criba.edu.ar} 
\and
Alberto Ibort\\
Departamento de Matem\'atica\\
Universidad Carlos III de Madrid\\
Av. de la Universidad 30,  Legan\'es, Madrid, Spain\\ 
{\footnotesize albertoi@math.uc3m.es}
\and
Manuel de Le\'on, \quad David Mart\'{\i}n de Diego\\
Instituto de Matem\'aticas y F{\'\i}sica Fundamental, CSIC\\
C/ Serrano 123, 28006 Madrid, Spain\\
{\footnotesize  mdeleon@imaff.cfmac.csic.es}\qquad
{\footnotesize d.martin@imaff.cfmac.csic.es}
}

\maketitle

\begin{abstract}
The constraint distribution in non-holonomic mechanics has a double role.  On one 
hand,
it is a kinematic constraint, that is, it is a restriction on the motion itself. 
On the
other hand, it is also a restriction on the allowed variations when using  
D'Alembert's
Principle to derive the equations of motion.   We will show that many systems of 
physical
interest where D'Alembert's Principle does not apply can be conveniently
modeled within
the general idea of the Principle of Virtual Work by the introduction of both 
kinematic constraints and variational constraints as being independent entities. 
This
includes, for example, elastic rolling bodies and pneumatic tires. Also, 
D'A\-lem\-bert's
Principle and Chetaev's Principle fall into this scheme.   We emphasize the 
geometric
point of view, avoiding the use of local coordinates, which is the appropriate 
setting
for dealing with questions of global nature, like reduction. 
\end{abstract}

\section{Introduction}
\paragraph{Non-holonomic Mechanics.}
The universal formalism created by La\-gran\-ge is not appropriate to derive the
equations of motion for systems with rolling constraints, that is, this motion is not described by classical variational principles. Several systems with  
rolling
constraints, like the idealized rigid ball rolling on a plane with only one point 
of
contact and many others,  are successfully described geometrically by 
distributions on
configuration space and the corresponding equations of motion are derived by
D'Alembert's Principle, which has been the purpose of extensive research
\cite{Bl96,CLMM,Ko92,LM1,LM,Marle,Fa72} for more than a century (see also for 
instance
\cite{Ne67,B03,CMR,Cort} for a list of references and historical remarks).  
However, as
we will see in the examples studied in the present work the dynamics of elastic
rolling bodies is not generally described by D'Alembert's principle, even in 
those
cases where the restriction on the motion is given by linear constraints.  On the 
other
hand,  {\bfseries\itshape second order constraints}, that is, subsets of the 
second
order tangent bundle rather than the tangent bundle of the configuration space, 
naturally appear in several examples. The purpose of the present work is to 
establish
the basic geometric definitions and procedures within the general idea of the
Principle of Virtual Work, generalizing D'Alembert's Principle to deal with 
nonlinear and higher order constraints. One of our main examples will be elastic rolling bodies, like
pneumatic tires, where some second order constraints appear naturally. 

In D'Alembert's principle the constraint distribution has a double role. 
On one hand, it is a {\bfseries\itshape kinematic constraint}, that is, it is a
restriction on the motion itself.  On the other hand, it is, {\em in addition}, a
{\bfseries\itshape variational constraint}.   This perspective was already adopted 
in
\cite{Ib96} where a general approach to non-holonomic constrained systems 
considered as
implicit differential equations was considered.  There it was discussed that the
kinematical constraints defining a submanifold on the tangent space of the
configuration space of the system and the reaction or control forces described by 
a subbundle of the cotangent bundle of the configuration space, were independent
entities and a condition for the compatibility of both ingredients was obtained.  
In this paper we will push forward this point of view by considering nonlinear higher order non-holonomic constraints, not only constraints on the velocities but on higher 
order derivatives.  

We will show that many systems of physical
interest where D'Alembert's Principle does not apply, can be conveniently modeled 
by a Principle based in the introduction of both higher order {\bfseries\itshape 
kinematic constraints} and higher order {\bfseries\itshape variational constraints} as 
being independent entities.   This includes, for example, elastic rolling bodies and 
pneumatic tires. Also, D'Alem\-bert's Principle and Chetaev's Principle fall into this 
scheme. 

Our point of view is geometric, avoiding the use of local coordinates, which  is
appropriate for dealing global problems, like reduction.  We also
write equations of motion for systems with higher order constraints in an 
intrinsic
fashion, using the natural structures of the tangent bundle and higher order 
bundles.   

\paragraph{Basic Notation}
As usual we will consider that the configuration space of a Lagrangian system is 
a smooth manifold $Q$ of dimension $n$ with local coordinates $q^i$.
We shall introduce higher order tangent bundles in order to deal with higher order
constraints.   Thus,  by definition, two given curves in
$Q$, say, $q_1(t)$ and $q_2(t)$, $t\in (-a, a),$
have a contact of order  $k$
at $q_0 = q_1(0) = q_2(0)$
if there is a local chart
$(\varphi, U)$
such that 
$q_i(0) \in U,$
for
$i = 1,2,$
and
$D_t^s \left(\varphi \circ q_1\right)(0) = D_t^s \left(\varphi \circ 
q_2\right)(0),$
for 
$s = 0,...,k.$
This is a well defined equivalence relation, and the equivalence class of a given 
curve
$q(t)$
is denoted
$[q]^{(k)}.$
For each
$q_0 \in Q,$
let
$T_{q_0}^{(k)} Q$
be the set of all
$[q]^{(k)}$
such that 
$q(0) = q_0,$
and let
$T^{(k)} Q$
be the collection of all
$T_{q_0}^{(k)} Q,$
for
$q_0 \in Q.$
It is well known (see for instance \cite{LR1}, \cite{Cr86} and references therein) 
that 
$\tau^k  : T^{(k)} Q \rightarrow Q,$
where
$\tau^k \left([q]^{(k)}\right) = q(0),$
is a fiber bundle, called the tangent bundle of order
$k$ of 
$Q.$
There are natural maps
$\tau^{(l,k)}  : T^{(k)} Q \rightarrow T^{(l)} Q,$
for
$k, l = 1,2,...,$
given by
$\tau^{(l,k)}\left([q]^{(k)}\right) = [q]^{(l)}.$
It is easy to see that
$T^{(1)} Q \equiv TQ.$
Also, we can identify
$T^{(0)} Q \equiv Q,$
via
$[q]^{(0)} \equiv q(0).$

In local coordinates, we have
$q = (q^1,...,q^n),$
and, for
$s = 1,2,..,$
we denote
$q^{(s)} = \left(q^{1,(s)},...,q^{n,(s)}\right),$
where
$$
q^{i,(s)} = \frac{d^s q^i}{dt^s}(0),
$$
where
$i = 1,...,n.$
Then we have,
$[q]^{(k)} = \left(q^{(0)},...,q^{(k)}\right).$

Denote by $j_k: T^{(k)}Q\rightarrow T(T^{(k-1)} Q)$ the canonical immersion 
defined by
$j_k([q]^{(k)})=[{q}^{(k-1)}]^{(1)}$ where ${q}^{(k-1)}$ is the lift to $T^{(k-
1)}Q$ of the curve $q$, that is,  the curve ${q}^{(k-1)}: (-a, a)\rightarrow 
T^{(k-1)}Q$ is defined as $q^{(k-1)}(t)=[q_t]^{(k-1)}$ where
$q_t(s)=q(t+s)$. 

In this paper, it will be useful to  introduce, geometrically, the concept of 
implicit
differential equations. This concept has often received less attention than the 
notion
of an explicit differential equation in the differential geometry literature (see
\cite{Ma92,Me95,Ib96}).  Geometrically, a system of implicit $k$th-order 
differential
equations is a submanifold $M$ of $T^{(k)}Q$ and a  curve $\gamma: 
I\longrightarrow Q$
is a solution to the differential equation $M$, if its $k$-lift 
${\gamma}^{(k)}(s)\in
M$ for all $s\in I$.  The implicit differential equation will be said to be 
integrable 
at a point if there exists a solution $\gamma$  such that  its $k$-lift passes 
through
it. The integrable part of $M$ is the subset of all integrable points of $M$. The
system is said to be integrable if  its integrable part coincides with  $M$.  A
notorious algorithm has been developed to extract the integrable part of an 
arbitrary
implicit differential equation \cite{Me95}, but it will not be the
objective of this paper to discuss this issue for 
systems with higher order non-holonomic constraints and we will restrict ourselves 
to
the description of the corresponding implicit differential equation, leaving the
questions of the existence and uniqueness of its solutions for future discussion.

\medskip

In section \ref{section2} we describe a first example of the elastic rolling 
ball, where some of the features of the general procedure already appear. In sections
\ref{section3} and \ref{section four} we show how to study Rocard's theory and also Greidanus's theory of a pneumatic tire (see \cite{Greidanus, Rocard1, Rocard} and also \cite{Ne67}) as a non-holonomic system with higher order constraints and, 
motivated by
the previous examples, in section \ref{section four} we establish a general principle for dealing with systems
involving higher order constraints.   The distinction between kinematic 
constraints and
variational constraints as being independent entities is a key point to this
discussion.  In Section \ref{section5} intrinsic equations of motion for systems with higher 
order constraints are derived.   In Section 6
further examples are provided and some basic results about reduction and the
equations for Lagrangian systems with symmetries with higher-order non-holonomic
constraints are discussed.   
\section{A Simple Example: the Elastic Homogeneous Rolling Ball}\label{section2}
The main purpose of this section is to show an example that can be treated using D'Alembert's Principle and also using some other procedures involving different types of constraints, including second order nonlinear constraints. All those procedures are equivalent in the sense that they give equivalent systems of equations.

Let us consider an elastic ball subjected to gravity and rolling on a plane. 
Without loss of generality we will assume that the radius of the ball is 1, for 
simplicity.
For a static ball the contact between the ball and the plane is a circle, whose
diameter was calculated by Hertz \cite{Hertz}, see also \cite{LL}, page 27. The effect of
internal viscosity, adhesion and other dissipative forces is important in some 
cases \cite{Br}, however, in the present section we shall assume that heat dissipation 
is small, in other words, we will consider only the idealized model of a perfect 
elastic ball. 
Also, we shall consider only  the important case where the circle of contact 
is small and the motion is quasistatic, which, in particular, implies that the zone 
of contact is approximately a circle of the same size as the contact circle in the
static case (see \cite{LL}). 
This also implies that the size and inertia of the
flattened part of the sphere is negligible.  Now we shall define the 
{\bfseries\itshape non-sliding condition}. 
It is given by the condition that the points of the sphere
belonging to the circle of contact cannot slide against the plane. It is clear 
that this has to be understood in an approximate sense since the exact solution of
elasticity equations is not known in general, not even  under the quasistatic
assumption. 
More precisely, we accept the following approximate model. We assume
that for all kinematic and dynamical purposes the ball is rigid, it has only one 
point of contact 
$a$
with the plane, representing the center of the circle of contact, which 
does not slides, and the spatial angular velocity and the translation velocity combine in such a way that the velocity of points 
$z$
of the surface of the ball near 
$a$
have a velocity which is of order
$|a-z|^2.$
This is 
a rigorous way of defining the constraint given by the non-sliding condition, in the 
case where there is a circle rather than a point of contact. It is easy to prove that, in fact, the non-sliding condition is satisfied if and only if the vertical component of the spatial angular velocity is 
$0,$
that is,
$\omega_3 = 0.$
We emphasize that this  model is
realistic only for slow motion and small deformation.
In agreement with all these physical assumptions we
have the following geometric model.

\paragraph{Kinematics of the Elastic Rolling Ball.}
The manifold
$Q = SO(3)\times \mathbb{R}^2$
is the configuration space for the model.
A position of the system is given by a point 
$(A, a) \in Q,$
where
$a$
is the point of contact of the sphere with the plane representing in the 
approximation described above the center of the circle of contact.
Let
$V = \dot{a}$
be the translation velocity of the ball
and let
$\omega = \dot{A} A^{-1}$
be the spatial angular velocity 
$\omega = (\omega_1, \omega_2, \omega_3),$ after the identification of $\mathfrak{so}(3)$ with
$\mathbb{R}^3$.
We have
$\dot{V} = \ddot{a}$
and
$\dot{\omega} = \ddot{A}A^{-1} - \dot{A}A^{-1}\dot{A}A^{-1}.$
The following two equations describe the non-sliding constraint
\begin{align}
V  &=  (\omega_2, -\omega_1)\label{constr1}\\
\label{constr2} \omega_3  &=  0.
\end{align}
The first equation represents the usual non-sliding condition for a rigid rolling 
ball while the second expresses the fact that there is really a circle of 
contact rather than a point, and that the points of that circle belonging to the 
sphere have zero velocity with respect to the plane, at 
least to first order approximation.  The previous equations define a distribution, 
which is the {\bfseries\itshape kinematic constraint} for the system of the 
elastic rolling ball.
We will  show that, provided that we accept higher order constraints, there are 
other equivalent ways of choosing the constraints all of them giving equivalent  
equations of motion.    For instance, let the curve
$a(t)$
in the plane have curvature radius
$r(t).$
Then we define the constraint
\begin{align}
r^2 \omega_3^2 
&= 
\omega_1^2 + \omega_2^2\label{constr3},
\end{align}
whose physical meaning is that the instantaneous motion of the sphere is a 
superposition of a rotation about some vertical axis, 
with angular velocity
$\omega_3,$
and the motion of rolling on the plane with speed
\begin{align}
|V| 
&= 
\sqrt{\omega_1^2 + \omega_2^2}\label{constr3'},
\end{align}
and the point of contact is located at a distance
$r$
from the vertical axis. 
This is an example of a
{\it second order constraint}, it is a {\bfseries\itshape kinematic constraint} in the terminology introduced in section \ref{section four}
and it is equivalent to the constraint
(\ref{constr2}), in the sense that it gives equivalent equations of motion, as we 
will explain  later. However, as we have said before the non-sliding condition is satisfied only if
$r = \infty,$
which of course implies
$\omega_3 = 0,$
or if
$\omega = 0.$
Equation (\ref{constr1}) has the following consequence
\[
\dot{V} = (\dot{\omega}_2, -\dot{\omega}_1).
\]
Let 
${\bf t}$
and
${\bf n}$
be the tangent and normal vectors to the curve
$a(t).$
We have
\[
|V|{\bf n} = \pm(\omega_1, \omega_2),
\]
and also 
\[
\dot{V} = \frac{d |V|}{dt}{\bf t} + \frac{|V|^2}{r} {\bf n}.
\]
Then we can deduce
\begin{align}
\langle|V|{\bf n}, \dot{V}\rangle
&=
\pm(\omega_1 \dot{\omega}_2 - \omega_2 \dot{\omega}_1)\\
&=
\frac{|V|^3}{r},
\end{align}
from which we obtain the constraint
(\ref{constr3}) in the form
\begin{align}\label{constr4}
\omega_1 \dot{\omega}_2 - \omega_2 \dot{\omega}_1
&=
\omega_3 (\omega_1^2 + \omega_2^2),
\end{align}
where the choice of the sign
$\pm$
is the only one consistent with the standard choice for the direction of the 
normal
${\bf n}$
and the sign of
$\omega_3$
for the given physical description.
We have a subset
$C \subseteq T^{(2)}Q,$ 
given by (\ref{constr1}) and (\ref{constr4}),
rewritten in terms of 
$\dot{a},$ $A,$ $\dot{A},$ and $\ddot{A}.$
This is a {\bfseries\itshape second order kinematic constraint}.
Observe that, in this example, the projection 
$\tau^{(1, 2)}_Q : T^{( 2)} Q \rightarrow TQ$
defines a distribution
$D \subseteq TQ$,
by
$D = \tau^{(1, 2)}_Q(C),$
which is given by (\ref{constr1}), and that rewritten in terms of
$A,$ $\dot{A},$ $a$ and $\dot{a}$, gives an expression linear in
$\dot{A}$ and $\dot{a}$.
\paragraph{Dynamics of the Elastic Rolling Ball.}

The Lagrangian is given by the kinetic energy
$$
L(A, a, \dot{A}, \dot{a})
=
\frac{1}{2} I (\dot{A}A^{-1})^2
+ 
\frac{1}{2}M (\dot{a})^2,
$$
where
$I$
is the moment of inertia of the ball with respect to any of its symmetry axis, and
$M$
is the mass of the ball.
The dynamics of the elastic rolling ball is given by the following variational 
description, as we will see later,
\begin{align}\label{VPW1}
\delta \int_{t_0}^{t_1}\left(\frac{1}{2} I (\dot{A}A^{-1})^2
\right.
&+ 
\left.
\frac{1}{2}M (\dot{a})^2\right)dt
= 0\\\label{rballdc1}
\left(\delta A(t_i), \delta a(t_i)\right)
&= 0, \quad\mbox{for}\quad i = 0,1\\\label{rballdc2}
\left(\delta A(t), \delta a(t)\right)
&
\in D_{\left(A(t), a(t)\right)}, \quad\mbox{for all}\quad t\\
\label{rballdck2}
\left(\dot{A}(t), \dot{a}(t)\right)
&
\in D_{\left(A(t), a(t)\right)}, \quad\mbox{for all}\quad t\\
\omega_3 = 0.
\end{align}
We will show that we can replace the last equation by equation (\ref{constr4})
and we will obtain an equivalent system.
We note that in this formulation the
{\bfseries\itshape constraints on the variations} are the same as in the case of the
rigid rolling ball (see for instance \cite{Ne67,B03}). However, the 
{\bfseries\itshape kinematic constraints} are not, in other words, the motion is effectively 
constrained by
our choice of the last equation, namely, either equation (\ref{constr2}) or 
equation (\ref{constr4}).  For any of those choices, we derive from the previous Principle a
{\bfseries\itshape differential-algebraic system} of equations and we will have 
existence and uniqueness of solution for those initial conditions compatible with the
constraints. 

By applying the usual integration by parts argument, we obtain the equations of 
motion.  However, as it already happens in the case of the rigid body, this is not 
completely trivial unless one is willing to use reduction arguments, 
(see for instance \cite{HLR} and \cite{CMR}).  We
will postpone the details of the computation until Section 6. 
We obtain,
\begin{align}\label{eqtnsmotion1}
(I + M)\dot{\omega}_1
&= 
0\\\label{eqtnsmotion2}
(I + M)\dot{\omega}_2
&= 
0\\\label{eqtnsmotion3}
(I + M)\dot{\omega}_3
&= 
0\\\label{eqtnsmotion4}
(\omega_2, - \omega_1) &=
V \\\label{eqtnsmotion5}
\omega_3 
&= 
0.
\end{align}
Of course this system is over determined, but it is correct.
The fifth equation, which coincides with equation
(\ref{constr2}),
may be replaced by
equation
(\ref{constr4}) and we obtain a system which is clearly equivalent.
The first four equations are exactly the equations for the rigid rolling ball and 
they imply that 
$\dot{\omega} = 0$
and also that the translation velocity
$V$
is constant.
We can show that there is solution provided that the initial 
condition
$(\omega_0, V_0)$
satisfies the constraints given by the last two equations and 
that this solution is unique.

We must remark at this point that the only guiding idea to establish the previous 
procedure is the Principle of Virtual Work, and one should check that the final 
equations are consistent with the  basic laws of mechanics, essentially Newton's 
Law, so the force should be equal to the rate of change of linear 
momentum and the torque should be equal to the rate of change of angular 
momentum.  In the case of the elastic rolling ball the forces of the constraint 
must satisfy the following conditions: the resultant force exerted by the plane on 
the ball has a positive component in the vertical upwards direction while the 
torque has a zero horizontal component. All this is obviously compatible with the 
previous system of equations. Moreover, the same equations can be derived by an 
elementary exercise in rational mechanics. We observe that preservation of energy is 
satisfied in this example. 
As a final remark to this example we observe that even if the constraints
(\ref{constr1}), (\ref{constr2}) are linear, we have not applied D'Alembert's Principle.
However, it will become clear at the end in section \ref{section4} that D'Alembert's Principle gives correct equations of motion in this example, and it is perhaps the best procedure in this case since it produces a non-overdetermined system. Showing that it is not always the case that D'Alembert's Principle can be applied is part of the purpose of the present work.
It is also clear from what we have explained so far that, for 
a given system, there is in principle the possibility of introducing several 
classes of higher order constraints which are equivalent in the sense that they 
lead to equivalent equations of motion.

The case of the nonhomogeneous elastic ball and also the case of the nonhomogeneous viscoelastic ball could be interesting, for instance because of possible applications to spherical robots, and can be treated with the methods of the present work. In particular, the non-sliding condition (\ref{constr2}) will be part of the kinematic constraints.
The case of the symmetric elastic or viscoelastic rolling ball, in which two of the three moments of inertia of the ball are equal, presents an extra symmetry and we can expect that some kind of reduction by this symmetry will help to understand the behavior of the reduced variables such as the angular momentum. The case of the rigid symmetric rolling ball has been studied in \cite{HLR}.

\section{An Example of Nonlinear Higher Order Non-holonomic Constraints}\label{section3}
In the example of the elastic rolling ball described in the previous section the 
second order constraint gives rise to a distribution
$D$
defined by (\ref{constr1})
which provides a restriction for the variations to obtain some of the equations of 
motion. The rest of the equations of motion are the ones given by the same 
distribution, plus an extra equation provided by the nonlinear second order constraint
(\ref{constr4}) or, equivalently, by the linear constraint (\ref{constr2}).  
This gives a procedure whose correctness in the example under consideration is 
established by the fundamental principles of mechanics. 
\paragraph{Rocard's Theory of a Pneumatic Tire.}
Before we try to establish any general procedure we will describe another 
example where the restrictions, both kinematic restrictions and restrictions on the variations, are 
of an entirely different nature. This is the simplified model of a pneumatic tire 
rolling on a plane according to Rocard's theory, as described for instance in 
\cite{Rocard}, \cite{Rocard1}, \cite{Ne67}.
For simplicity we shall study the case of a single elastic pneumatic tire whose 
plane is constrained to remain vertical while it rolls without sliding. 
The zone of contact of the pneumatic tire with the plane is a small surface with a 
central point of contact 
$x = (x_1, x_2),$
which for simplicity we will assume that it coincides with the projection of the center of the wheel on the plane.
The non-sliding condition means that the velocity of the points of  the tire 
belonging to the zone of contact with respect to the plane is zero. In an approximate sense this non-sliding condition implies that the vertical component 
of the angular velocity of the small piece of surface of the pneumatic in contact 
with the floor is zero. 
However, contrary to what we have assumed for the homogeneous elastic rolling ball, the fact that the vertical component of the angular velocity of the zone of contact is 
zero does not mean that the vertical component of the angular velocity of the 
plane of the tire is zero. This is 
because according to Rocard's theory the elasticity of the material allows for a 
small angle
$\epsilon$
between the axis of the zone of contact(an oblong-like symmetric zone), which is assumed to have the direction of
$\dot{x},$
and the plane of the wheel.
We will call 
$K$
the corresponding constant of elasticity.
It turns out that the non-sliding condition for the small zone of contact is not the relevant constraint. Instead, there will appear another second order constraint of a 
different nature. Finally, we must remark that the previous description of Rocard's theory gives only an approximation, and for more accurate results one must have into account some other observed effects. For instance, the projection 
$x$
of the center of the wheel onto the plane is not exactly the center of the zone of contact, which produce a small torque not taken into account in the simplified model described above. Part, but only part, of this problem is taken into account in the simplified version of Greidanu's theory described later in the present work.

Taking into account all the physical considerations explained above we will 
describe Rocard's theory  by the following geometric model.
For all kinematic and dynamical purposes the wheel is simply an undeformable 
disk kept vertical and rolling on a plane, where the point of contact is
$x = (x_1, x_2).$
We choose once for all a normal vector 
$N = (-\operatorname{sin}\theta , \operatorname{cos}\theta )$
rigidly fixed to the wheel.
Then the angle between the plane of the wheel and the
$x_1$
axis is
$\theta.$
The angle between the velocity vector 
$\dot{x}$ 
and the plane of the wheel is called
$\epsilon,$
with the physical meaning that we have explained before.
Therefore, the angle between the axis
$x_1$
and
$\dot{x}$
is
$\theta - \epsilon,$
and the vector 
${\bf n},$
normal to the trajectory of the point
$x$
and pointing in the direction of the concavity of the curve, is
${\bf n} = \left(-\operatorname{sin}(\theta - \epsilon) , 
\operatorname{cos}(\theta - \epsilon) \right).$
The angle of rotation of the wheel about its own axis is called
$\psi.$
In order to obtain precise formulas one should be careful about the sign conventions.
Positive angles in the 
$x_1 x_2$
plane satisfy the usual convention. Thus the angle between the
$x_1$
axis and the
$x_2$
axis is, by definition, 
$(1/2)\pi$
while the angle between the
$x_2$
axis and the
$x_1$
axis is 
$-(1/2)\pi.$
The sign for the angle 
$\psi$
is established by the convention that the vector angular velocity is of the form
$\dot{\psi} N.$
The configuration space of the system is
$Q = \mathbb{T}^3 \times \mathbb{R}^2,$
and a generic point is
$q = (q_1, q_2, q_3, q_4, q_5)\equiv (\psi, \theta, \epsilon, x_1, x_2).$
The Lagrangian is given by
\[
L(q, \dot{q})
=
\frac{1}{2} I \dot{\psi}^2 
+
\frac{1}{2} J \dot{\theta}^2 
+
\frac{1}{2}M\dot{x}^2 
-
\frac{1}{2}K\epsilon^2,
\]
where
$I$
is the moment of inertia of the wheel with respect to its axis, 
$J$
is the moment of inertia of the wheel with respect to any one of its diameters,
$M$
is the mass of the wheel and
$K$
is the constant of elasticity introduced before, which by definition satisfies
$T = -K\epsilon,$
where 
$T$
is the vertical torque. The kinetic energy due to the velocity of rotation 
$\dot{\epsilon}$
of the small flattened piece of material about the zone of contact is 
small and we will assume that it is
$0$
for simplicity, which is also in agreement with general standard assumptions for this kind of approximate models, \cite{Ne67}.

Next we shall describe the kinematic constraints and the variational constraints.
The {kinematic constraint} 
$C_K,$
is given by the equations
\begin{align}\label{kconst1}
\dot{x}_1 
&=
\dot{\psi}\operatorname{cos}(\theta - \epsilon)\\\label{kconst2}
\dot{x}_2 
&=
\dot{\psi}\operatorname{sin}(\theta - \epsilon)\\\label{kconst3}
-\ddot{\psi}\operatorname{tg}\epsilon
+
\dot{\psi}(\dot{\theta} - \dot{\epsilon})
&=
(\operatorname{sign}{\dot{\psi}})\frac{a}{M}\operatorname{tg}\epsilon .
\end{align}
The first two equations represent the non-sliding condition for the center of the zone of 
contact, and they are the same as the ones that appear in the case of a rigid 
rolling disk, or wheel, except for the small angle
$\epsilon.$
We should emphasize that here we are working to first order approximation only, 
which means that powers of
$\epsilon$
greater than 
$1$
may be neglected. 
The last equation comes from Rocard's condition,
\[
|F| = a\operatorname{sin}|\epsilon|,
\]
where
$a$
is a positive physical constant and
$F$
is the force normal to the wheel exerted by the floor, while the wheel is rolling 
with
nonzero velocity. More precisely, 
$F$
is the $N$ component of the centripetal force, 
that is we have
$F = <M\ddot{x},\, N>.$
The sign conventions are encoded in the following more precise version of Rocard's formula
\[
F = (\operatorname{sign}\dot{\psi})a\operatorname{sin}\epsilon,
\]
where
$\epsilon$
must be interpreted as being the angle between the normal 
$\textbf{n}$
to the curve and 
$N$
if
$F>0$
while it must be interpreted as being the angle between 
$\textbf{n}$
and
$-N$
if
$F<0.$
Recall that Rocards's formula is valid for
$\epsilon$
close to
$0$
only.
A couple of remarks is in order for future use.
First, as we have said before, Rocard's theory is valid modulo infinitesimals of order
$(\operatorname{sin}\epsilon)^2.$
Second, with the previous sign conventions and according to Rocard's formula it is not difficult to show that
$\epsilon(\dot{\theta} - \dot{\epsilon})\geq 0.$
It also follows from the expression of Rocard's formula given by
(\ref{kconst3}) that for 
$\epsilon = 0$
the curve
$x(t)$
must have a point of inflection, that is 
$\dot{\theta} - \dot{\epsilon} = 0.$

It is clear that (\ref{kconst3}) involves the first and second derivatives of some 
of the variables with respect to time, moreover, the dependence on the first 
derivatives is nonlinear, therefore it is far from the typical constraints of 
D'Alembert type. 
To obtain equation
(\ref{kconst3})
we may assume, without loss of generality, that
$\dot{\psi} > 0.$
We simply differentiate 
(\ref{kconst1}) and (\ref{kconst2})
with respect to time, and replace in the equation
$(\operatorname{sign}\dot{\psi})a\operatorname{sin}\epsilon = <M\ddot{x}, N>.$
Now let us consider the following
{variational constraints} 
$C_V,$
to be imposed on variations
$\delta{q}$
\begin{align}\label{dconst1}
\delta{\psi}\operatorname{cos}\theta
-
\delta{x}_1 
&= 0\\\label{dconst2}
\delta{\psi}\operatorname{sin}\theta
-
\delta{x}_2
&=
0\\\label{dconst3}
\delta{\theta} - \delta{\epsilon}
&=
0.
\end{align}
Consider the curves
$q(t)$
satisfying
\[
\delta \int_{t_0}^{t_1} L(q, \dot{q})dt = 0,
\]
for variations 
$\delta q$
satisfying
$\delta q (t_i) = 0,$
for
$i = 1,2,$
and also the variational constraints
$C_V.$
Those curves are the ones satisfying the following 
{\bfseries\itshape dynamic equations}
\begin{align}\label{dyneq1}
I\ddot{\psi}
+
M\ddot{x}_1 \operatorname{cos}\theta
+
M\ddot{x}_2 \operatorname{sin}\theta
&=
0\\\label{dyneq2}
J\ddot{\theta} + K\epsilon 
&=
0,
\end{align}
obtained by the usual integration by parts arguments.
These dynamic equations give balance between forces of the constraint and rate of 
change of momentum. The resultant of the forces exerted by the plane of contact on 
the wheel 
has positive upwards vertical component which is compensated by gravity, while 
the horizontal component, which is given by
$M\ddot{x},$
is decomposed in the directions
$(\operatorname{cos}\theta, \operatorname{sin}\theta)$
and
$(-\operatorname{sin}\theta, \operatorname{cos}\theta).$
The first one is compensated by the rate of change of the angular momentum
$I\ddot{\psi}$
and the second is compensated by the non-sliding constraint force.
The vertical component of the torque of the forces exerted by the plane on the 
wheel is 
$K\epsilon$ 
which is compensated by
$J\ddot{\theta}.$
The other components of the torque are automatically compensated because we are 
assuming that the wheel is forced to remain vertical.
The system of dynamic equations together with the kinematic constraints equations
$C_K$
completely describe the motion of the wheel.

\medskip

In the previous example, we should emphasize, again, the distinction between 
{\bfseries\itshape kinematic constraints} and {\bfseries\itshape variational 
constraints}. They are conceptually different, and this difference is implicit in 
the usual statement of the Principle of Virtual Work. However, in the literature this 
distinction is usually not emphasized, and for good reason, since in those cases where
D'Alembert's principle can be applied 
the variational constraints and the kinematic constraints coincide. 
Non-holonomic systems that cannot be treated using D'Alembert method have been 
considered for instance by Chetaev \cite{Ch32} where a procedure to deal with 
general first order nonlinear constraints is devised (see also \cite{Ap11,Pi82}). 
In Marle \cite{Marle} it is clearly stated that constraint forces cannot be 
derived in general from the kinematic constraints and have to be added as part of the 
physical description of the system.   Furthermore in \cite{Ib96} it was explicitly
stated a formulation for first order Lagrangian and Poisson nonholonomic systems where
kinematic constraints and constraint forces are given as independent entities.

In the case of the elastic rolling ball the forces of the constraint are normal to the direction of the motion of the ball and there is no dissipation of energy. However, for a viscoelastic rolling ball there is certainly dissipation of energy and the component of the force of the constraint in the direction of the motion can be calculated using results from \cite{Br}. This kind of system can also be approached using the kind of generalization of D'Alembert's principle described in section \ref{section four}. 
The rate of dissipation of energy for a pneumatic tire rolling according to Rocard's theory can be easily calculated. Since the energy is given by
$E = (1/2) I \dot{\psi}^2 
+
(1/2) J \dot{\theta}^2 
+
(1/2) M\dot{x}^2 
+
(1/2) K\epsilon^2,$
using the kinematic constraints 
(\ref{kconst1}), (\ref{kconst2})
and the dynamic equations derived before we can show after some easy calculations that
$\dot{E} = -\left(M\dot{\psi}^2 + K\right)\epsilon (\dot{\theta} - \dot{\epsilon}),$
modulo infinitesimals of order
$\epsilon^2.$
Since
$\epsilon (\dot{\theta} - \dot{\epsilon}) \geq 0$
as we have explained before we have
$\dot{E}\leqq 0,$
which means that in general there is dissipation of energy. The limit case
$\epsilon = 0$
gives
$\dot{E} = 0,$
which reveals that Rocard's theory does not take into account the relatively small dissipation of energy that occurs when the tire rolls in a straight line.
To prove the previous formula we proceed as follows.
We can easily see that
$\dot{E} 
= 
I\dot{\psi}\ddot{\psi}
+ 
J\dot{\theta}\ddot{\theta}
+
M\dot{x}\cdot \ddot{x}
+
K \epsilon \dot{\epsilon}.$
By differentiating (\ref{kconst1}) and (\ref{kconst2}) we can easily see that
$\dot{x}\cdot \ddot{x} = \dot{\psi}\ddot{\psi}$
and from this and the dynamic equation (\ref{dyneq2}) we obtain
$(I + M)\dot{\psi}\ddot{\psi} - K\epsilon (\dot{\theta} - \dot{\epsilon}) = 0.$
Using (\ref{kconst1}), (\ref{kconst2}) and (\ref{dyneq1}) we obtain, modulo higher order infinitesimals,
that
$(I + M)\ddot{\psi} = -M\dot{\psi}\epsilon (\dot{\theta} - \dot{\epsilon})$
therefore
$(I + M)\ddot{\psi}\dot{\psi} = -M\dot{\psi}^2\epsilon (\dot{\theta} - \dot{\epsilon}),$
from which we finally obtain
$\dot{E} = -\left(M\dot{\psi}^2 + K\right)\epsilon (\dot{\theta} - \dot{\epsilon}).$
>From a general point of view we may say that the distinction between variational and kinematic constraints implies that the infinitesimal work of the constraint forces in general does not vanish for some admissible infinitesimal displacements, which is the reason why the forces of the constraint may produce work.

In the next section we define a class of non-holonomic systems with higher order 
nonlinear constraints based on the introduction of both kinematic  and 
variational constraints. We will also show that procedures like D'Alembert's Principle or 
Chetaev's procedure fall into this scheme. 
We propose that questions of a general nature on non-holonomic systems, like 
reduction by the symmetry,  Legendre transformation, and many others should be 
approached for the general case of higher order constraints using the scheme 
based on the introduction of both kinematic  and variational constraints. 
\medskip

\section{A Principle of Virtual Work for Lagrangian Systems with Nonlinear Higher order
Non-holonomic Constraints}\label{section four}

Let $Q$ be a configuration space of dimension
$n$ and let
$L \colon TQ \rightarrow \mathbb{R}$
be a given Lagrangian. Then we have the Euler-Lagrange operator
$\mathcal{EL} : T^{(2)} Q \rightarrow T^\ast Q$
which is given in coordinates by
\[
\mathcal{EL}_i([q]^{(2)})\delta q^i
=
\left(\frac{d}{dt}\frac{\partial L}{\partial \dot{q}^i}\left([q]^{(2)}\right)
-
\frac{\partial L}{\partial q}\left([q]^{(2)}\right)\right)
\delta q^i.
\]
A {\bfseries\itshape kinematic constraint of order} $k$ is, by definition, a
subset 
$C_K \subseteq T^{(k)}Q,$
for some
$k = 0, 1, 2,...$
The subset
$C_K$
is often defined by equations
$R_K \left([q]^{(k)}\right) = 0,$
where
$R_K : T^{(k)}Q \rightarrow \mathbb{R}^r,$
for some
$r = 1, 2,....$
For example if 
$k = 0$
and 
$R_K$
is a submersion then
$C_K$
is a nonsingular holonomic constraint. If 
$k = 1$
and 
$R_K (q, \dot{q}) = R_{Ki} (q)\dot{q}^i$
defines a distribution of constant rank, we have the typical situation of 
D'Alembert's Principle.
If
$R_K (q, \dot{q})$
is a general function we have the situation considered by Chetaev \cite{Ch32}. In 
the case of the elastic rolling ball we have,
if we choose the constraint given by equation
(\ref{constr2})
as we have explained before,
$n = 5,$
$k = 1,$ 
$r = 3,$
and
$$
R_K (A, a, \dot{A}, \dot{a}) 
= 
(\omega_2 - \dot{a}_1, -\omega_1 - \dot{a}_2, \omega_3).
$$
Alternatively, as we have explained before, if we choose the constraint given by 
equation
(\ref{constr4}),
we have,
$n = 5,$
$k = 2,$ 
$r = 3,$
$$
R_K (A, a, \dot{A}, \dot{a}, \ddot{A}, \ddot{a}) 
= 
(\omega_2 - \dot{a}_1 , -\omega_1 - \dot{a}_2, \omega_1 \dot{\omega}_2 - \omega_2 
\dot{\omega}_1
-
\omega_3 (\omega_1^2 + \omega_2^2)).
$$
In the case of the Rocard's theory of a pneumatic tire, we have
$n = 5,$
$k = 2,$
$r = 3,$
and
\begin{align}\label{}
R_K
&
\left(\psi, \theta,  \epsilon, x_1, x_2,
\dot{\psi}, \dot{\theta}, \dot{\epsilon}, \dot{x_1}, \dot{x_2}\right)\\
&=
\left(\dot{x}_1 
-
\dot{\psi}\operatorname{cos}(\theta - \epsilon), 
\dot{x}_2
-
\dot{\psi}\operatorname{sin}(\theta - \epsilon),
-\ddot{\psi}\operatorname{tg}\epsilon
+
\dot{\psi}(\dot{\theta} - \dot{\epsilon})
-
(\operatorname{sign}\dot{\psi})\frac{a}{M}\operatorname{tg}\epsilon \right).
\end{align}

A {\bfseries\itshape constraint on the variations of order} $l$ is a subset
$C_V \subseteq T^{(l)}Q \times_{Q} TQ$
defined by equations
$R_V \left([q]^{(l)}, \delta q\right) = 0$
where
$R_V$ 
is linear in the variable
$\delta q,$
so we shall write as usual 
$R_V \left([q]^{(l)}, \delta q\right)
=
R_V \left([q]^{(l)}\right).\delta q$
or, in coordinates,
$R_V \left([q]^{(l)}, \delta q\right) 
=
R_{Vi} \left([q]^{(l)}\right)\cdot \delta q^i.$
For each
$[q]^{(l)}\in T^{(l)}Q,$
we let
$C_V\left([q]^{(l)}\right)
=
\{\delta q \in TQ: \left([q]^{(l)}, \delta q \right)
\in C_V\}.$
\medskip

\paragraph{Statement of the Principle.}The main object defined in this paper is the class of Lagrangian non-holonomic systems 
defined by data
$(L, C_K, C_V)$
whose {\bfseries\itshape dynamical equations} are  
derived by using the variational principle
\[
\delta \int_{t_0}^{t_1} L(q, \dot{q})dt = 0,
\]
where variations $\delta q$
are restricted by $\delta q \in  C_V\left([q]^{(l)}\right)$,
or, equivalently,
$R_V \left([q]^{(l)}\right)\cdot \delta q = 0.$
Then the {\bfseries\itshape equations of motion} are given by the {\bfseries\itshape 
dynamical equations}
\[
\mathcal{EL}_i([q]^{(2)}) \in  R_V\left([q]^{(l)}\right)^o 
\]
and the {\bfseries\itshape kinematic constraint} equations
$[q]^{(l)} \in C_K$ or, equivalently,
$$R_K \left([q]^{(k)}\right) = 0.$$

\noindent Equations of motion will be derived in the next section.
\medskip

The previous Principle, which is contained in the general idea of the Principle of Virtual Work, imposses, through the dynamical equations, restrictions on the forces of the constraints. But, contrary to what happens with D'Alembert's Principle, the forces of the constraints derived from the Principle stated above will in general produce work.
\smallskip

The class of higher order non-holonomic systems just defined contains several
important classes of non-holonomic systems.   For example, for the class of 
non-holonomic systems that are tractable using D'Alembert's principle we have, by definition,
$k = 1,$
$l = 0$
and
$C_K$
is the distribution where for each
$q \in Q$
the space of the distribution is
$C_V(q)\subseteq TQ.$
Thus, the kinematic constraint and the constraint on the variations essentially
coincide in this case. In the case of nonlinear kinematic constraints considered 
by
Chetaev given by
$R_K(q, \dot{q}) = 0$
we have
$l = 1$ 
and the variational constraints are defined, according to Chetaev, by
\[
R_V (q, \dot{q})\cdot \delta q 
=
\frac{\partial R_K(q, \dot{q})}{\partial \dot{q}}
\cdot \delta q.
\]

\begin{remark}
{\rm 
In the mathematical literature one finds some  examples of higher order 
constraints in
non-holonomic problems (for instance see \cite{DS,Pi82,Krup,Va58}). In the 
previous references an extension of the Chetaev principle for kinematic second order 
constraints is applied, namely,  
\[
(R_{K})_i(q, \dot{q}, \ddot{q})=0,\  1\leq i\leq m
\]
and  variational constraints $R_V$ are derived from the kinematic constraints by 
\[
R_V(q, \dot{q}, \ddot{q})\cdot \delta q=\frac{\partial R_K}{\partial 
\ddot{q}}\cdot \delta q=0
\]
}
\end{remark}

In the case of the elastic rolling ball the variational constraints are given by 
(\ref{rballdc2}).
In the case of the pneumatic tire according to Rocard's theory the kinematic
constraints are given by  (\ref{kconst1}), (\ref{kconst2}), (\ref{kconst3}) and 
the variational constraints are given by 
(\ref{dconst1}),
(\ref{dconst2}), (\ref{dconst3}).

We emphasize once again that the notions of kinematic constraints and 
variational constraints are independent and one should not attempt, for instance, to derive variational constraints from kinematic constraints by a universal procedure.  In 
order to illustrate further the necessity of such a point of view we will describe next 
the example of Greidanus's theory of a pneumatic tire, where the kinematic constraint 
defines a distribution like in D'Alembert's Principle but the variational constraints 
are not given by the same distribution, therefore they are not the ones prescribed by
D'Alembert's Principle.
\paragraph{Pneumatic tires according to Greidanus}
Several approaches to the dynamics of a pneumatic tire like  those of Rocard,
Greidanus, Keldys and others can be found in \cite{Greidanus}, \cite{Rocard}, \cite{Rocard1}, \cite{Ne67}. To describe Greidanu's
approach we shall consider the simpler setting of Rocard's approach described 
before,
but this time we allow, in addition, for a lateral deformation
$\xi.$ The absolute value of the quantity
$\xi$
is the distance between the projection of the center of the wheel on the plane
$(x_1, x_2)$
and the center of the zone of contact. In the Rocard's approach described above the value of 
$\xi$
is 
$0.$
We must remark that we are considering in this paper only the case of Greidanus's theory in which the wheel is kept vertical. The physical reason for the appearance of the displacement
$\xi$
is of course the lateral deformation due to the centrifugal force given the elasticity of the material.

The kinematic constraints are
\begin{align}\label{kgre1}
\dot{x}_1 
&=
\dot{\psi}\operatorname{cos}(\theta - \epsilon)\\\label{kgre2}
\dot{x}_2 
&=
\dot{\psi}\operatorname{sin}(\theta - \epsilon)\\\label{kgre3}
\dot{\theta} - \dot{\epsilon}
&=
\dot{\psi}(\alpha \xi + \beta \epsilon).
\end{align}
The first two equations are the same as in Rocard's approach. The last one 
expresses
the fact that the curvature of the trajectory of the center of the contact zone 
is,
for a given speed of rotation of the wheel, proportional to a linear combination of the deformation parameters
$\xi$
and
$\epsilon,$
where
$\alpha > 0$
and
$\beta >0.$
This replaces Rocard's constraint.
We see that the kinematic constraints define a distribution. 
The variational constraints are 
\begin{align}\label{}
\delta x_1 
&=
\delta \psi \operatorname{cos}\theta\\
\delta x_2 
&=
\delta \psi \operatorname{sin}\theta\\
\delta \theta - \delta \epsilon 
&= 0.
\end{align}
These variational constraints are different from the kinematic constraints, 
therefore we are not using here D'Alembert's Principle.
The projection of the center of the wheel on the plane is the point
$(y_1, y_2)$
given by
\begin{align}\label{}
y_1 
&=
x_1 + \xi \operatorname{sin}\theta\\
y_2 
&=
x_2 - \xi \operatorname{cos}\theta.
\end{align}
It is more convenient to calculate the kinematic constraints and the variational 
constraints in terms of
$y_1$
and 
$y_2$
instead of
$x_1$
and 
$x_2.$
The kinematic constraints are
\begin{align}\label{kcons4}
\dot{y}_1
&=
\dot{\psi}\operatorname{cos}(\theta - \epsilon)
+
\dot{\xi}\operatorname{sin}\theta
+
\xi (\operatorname{cos}\theta )\dot{\theta}\\\label{kcons5}
\dot{y}_2
&=
\dot{\psi}\operatorname{sin}(\theta - \epsilon)
-
\dot{\xi}\operatorname{cos}\theta
+
\xi (\operatorname{sin}\theta )\dot{\theta}\\\label{kcons6}
\dot{\theta} - \dot{\epsilon}
&=
\dot{\psi}(\alpha \xi + \beta \epsilon).
\end{align}
The variational constraints are
\begin{align}\label{}
\delta y_1
&=
\delta \psi\operatorname{cos}\theta
+
\delta \xi\operatorname{sin}\theta
+
\xi (\operatorname{cos}\theta )\delta \theta\\
\delta y_2
&=
\delta \psi\operatorname{sin}\theta
-
\delta \xi \operatorname{cos}\theta
+
\xi (\operatorname{sin}\theta )\delta \theta\\
\delta \theta - \delta \epsilon
&=
0.
\end{align}
The Lagrangian is
\begin{eqnarray*}
L(\psi, \theta, \epsilon, y_1, y_2, \xi, 
\dot{\psi}, \dot{\theta}, \dot{\epsilon}, \dot{y}_1, \dot{y}_2, \dot{\xi} )
&=&
\frac{1}{2}I\dot{\psi}^2
+
\frac{1}{2}J\dot{\theta}^2
\\
&&+
\frac{1}{2}M\left((\dot{y}_1)^2 + (\dot{y}_2)^2\right)
-
\frac{1}{2}\alpha\xi^2 
-
\frac{1}{2}\beta\epsilon^2.
\end{eqnarray*}
Then, equations of motion are given by kinematic constraints
(\ref{kcons4}), (\ref{kcons5}), (\ref{kcons6}) and dynamic equations
\begin{align}\label{gdyn1}
I\ddot{\psi}
+
M\ddot{y}_1 \operatorname{cos}\theta
+
M\ddot{y}_2 \operatorname{sin}\theta
&= 0\\\label{gdyn2}
J\ddot{\theta}
+
M\xi\ddot{y}_1 \operatorname{cos}\theta
+
M\xi\ddot{y}_2 \operatorname{sin}\theta + \beta\epsilon
&= 0\\\label{gdyn3}
-
M\ddot{y}_1 \operatorname{sin}\theta
+
M\ddot{y}_2 \operatorname{cos}\theta - \alpha\xi
&= 0.
\end{align}
We can easily check that the previous equations represent the balance between rate 
of change of momentum and forces of the constraints.

For high values of
$\alpha$
the deformation
$\xi$
remains small. Moreover, for
$\alpha \rightarrow \infty$
we have
$\xi \rightarrow 0$
and the dynamic equations (\ref{gdyn1}), (\ref{gdyn2}) of Greidanu's theory become the equations (\ref{dyneq1}), (\ref{dyneq2}) of Rocard's theory, provided that
$K = \beta .$ Using this and the fact that the two first kinematic
constraints (\ref{kconst1}), (\ref{kconst2}) of Rocard's theory coincide with the first two kinematic constraints (\ref{kgre1}), (\ref{kgre2}) of Greidanu's theory and also the fact that for 
$\alpha \rightarrow \infty$
the mechanical energy
$E$
for both theories tend to the same value,
one can prove, proceeding as in the case of Rocard's theory, that at least for high values of 
$\alpha$
a pneumatic tire moving according to Greidanus theory is a dissipative system. This shows that D'Alembert's Principle does not provides a good model for this kind of system., even though the kinematic constraints are linear.
\section{Equations of motion}\label{section5}
Let us recall some basic facts of the geometry of the tangent bundle. The 
{\bfseries\itshape vertical endomorphism} $S$ is defined in local natural 
coordinates
$(q^A, \dot{q}^A)$ on $TQ$ by
$$ S = \frac{\partial}{\partial \dot{q}^A} \otimes d\, q^A\; .$$ 
The Liouville vector field $\Delta$ on $TQ$ is locally defined by
\[
\Delta=\dot{q}^A\frac{\partial}{\partial \dot{q}^A}\; .
\]
A second order differential equation is a vector field  $\Gamma$ on $TQ$ such that 
$S (\Gamma ) = \Delta$. 
We have the following local expression for $\Gamma$:
$$
\Gamma = \dot{q}^A \frac{\partial}{\partial q^A} + F^A(q, \dot{q}) 
\frac{\partial}{\partial  \dot{q}^A}\;  .
$$
An integral curve of $\Gamma$ is always the tangent prolongation
of its projection $q(t)$ on $Q$, called a {\bfseries\itshape solution} of 
$\Gamma$. It satisfies the following explicit system of second order differential 
equations:
$$
\frac{d^2q^A}{dt^2}=F^A(q, \dot{q})\;  .
$$
We also note that the kernel and image of
$S$ consist of vertical vector fields.   Moreover,  $S$ acts by duality
on forms and the kernel and image of $S^*$ consists of horizontal 1--forms.

Given a lagrangian function $L: TQ\longrightarrow \mathbb{R}$, we
construct the two-form $\omega_L = - d(S^* (dL))$ on $TQ$, and
the energy function $E_L = \Delta L - L$ (see \cite{LR2}).
A remarkable property of $S$ and $\omega_L$ is the following 
$ i_S \omega_L = 0 ,$
or, in other words,
\begin{equation}\label{S_omega} 
S^* \circ \hat{\omega}_L = - \hat{\omega}_L \circ S ,
\end{equation}
where $\hat{\omega}_L$ denotes the map $T(TQ) \to T^* (TQ)$ defined by
contraction with $\omega_L$.

Observe that if  $L$ is regular, then $\omega_L$ is a symplectic form, and 
there is a unique vector field $\Gamma_L$ satisfying
$$
i_{\Gamma_L} \, \omega_L = dE_L,
$$
or, in other words, $\Gamma_L$ is the Hamiltonian vector field
with Hamiltonian energy $E_L$. It is  well known that $\Gamma_L$ is a second order 
differential equation on $TQ$,
namely, the Euler-Lagrange equations for $L$.

Without the regularity condition, the Euler-Lagrange equations form  a  system of 
second order differential equations in $Q$, in implicit form, that is, a 
submanifold $D_2$ of $T^{(2)}Q$, determined by: 
\begin{equation}\label{asd}
D_2=\{w\in T^{(2)} Q\; |\; i_{j_2(w)}\omega_L(\tau^{(1,2)}(w))=dE_L( 
\tau^{(1,2)}(w))\}
\end{equation}  
or, in other words, 
\[
D_2=\{ w\in T^{(2)} Q\; |\; \mathcal{EL}(w)=0\}\; .
\]
The class of higher order non-holonomic systems  studied in this paper,  are 
determined by  data $(L, C_K, C_V)$. Next we will show that   the equations of 
motion of this kind of systems is a system of implicit $k$th-order differential 
equations. In what follows, and without loss of generality, we will always suppose 
that $k\geq l$ and $k\geq 2$.

In our case the constraint on the variations are determined by a subset 
$C_V\subseteq T^{(l)}Q\times _Q TQ$. Therefore for each point $[q]^{(l)}$ we 
obtain the annihilator $C^0_V([q]^{(l)})\subseteq T_{q}^*Q$ of $C_V([q]^{(l)})$.
Denote by $F_V([q]^{(l)})$ the subspace of $T^*(TQ)$ determined by 
$F_V([q]^{(l)})=(\tau_Q)^*(C^0_V([q]^{(l)}))$. 
Now, we shall define the subset of $T^{(k)}Q$: 
\[
M_V=\{ [q]^{(k)}\in T^{k}Q\; |\; i_{j_2([q]^{(2)})}\omega_L([q]^{(1)})-
dE_L([q]^{(1)})\in F_V([q]^{(l)})\}\; .
\]
Therefore, the non-holonomic  system associated to $(L, C_K, C_V)$, determines a 
$k$th-order  implicit system 
 given by the submanifold $M_{KV}=C_K\cap M_V$. The solutions of the problem $(L, 
C_K, C_V)$ are the  curves $\map{\gamma}{I}{Q}$ such that $\gamma^{(k)}\subset 
M_{KV}$.

\section{Further Results and Examples}\label{section4}
The scheme generalizing D'Alembert Principle, for the case of higher order 
constraints described in section \ref{section four} is not of course the most 
general case. It is not the purpose of the present work to expose the most general 
possible formalism, but on the contrary, to provide a scheme which is useful in a 
variety of problems in mechanics. 
This scheme is also useful to deal with important questions of a general character in mechanics, like reduction, Legendre transformation and 
others.
Some of these questions will be the purpose of future work and in this section 
we will consider some partial results only.
\paragraph{Reduction of Invariant Systems with Higher Order Constraints on a 
Group.}
In this paragraph we explain how to reduce invariant Lagrangian systems with 
higher order non-holonomic constraints on a group. The more general case of systems on a
principal bundle will be the purpose of a future work. However, in the present 
section we will show how to proceed in an example where the bundle is 
trivial,
which illustrates some of the features of the general theory. Assume that the
configuration space is a group
$G$
and that the Lagrangian
$L,$
the kinematic constraint
$C_K$
and the constraint on the variations
$C_V$
are left invariant. For right invariant systems we can proceed in a similar way.
For each 
$r = 1, 2,...$
we have an identification
$$
\alpha_r : T^{(r)}G/G \rightarrow r\mathfrak{g},
$$
where
$r\mathfrak{g} = \mathfrak{g}\oplus...\oplus\mathfrak{g},$
is the direct sum of 
$r$
copies of
$\mathfrak{g}.$
This identification is uniquely defined by the map
$[g]^{(r)} \rightarrow [v]^{(r)},$
where
$v = g^{-1}\dot{g},$
and
$[v]^{(r)} = \left(v^{(0)}, v^{(1)},...v^{(r-1)}\right),$
where, by definition, we have,
$$
v^{(i)} 
=
\frac{d^{i}}{dt^{i}}v,
$$
for 
$r = 0,1,...r-1.$
Under the identification
$\alpha_k,$
the quotient of the kinematic constraint
$C_K/G,$
becomes a subset, called {\bfseries\itshape reduced kinematic constraint},
$\mathfrak{C}_K \subseteq k\mathfrak{g}.$
Similarly, for each 
$r = 1,2,...$
we have an identification
$$
\beta_r : \left(T^{(r)}G \times_{G} TG\right)/G \rightarrow r\mathfrak{g}\oplus 
\mathfrak{g},
$$
This identification is uniquely defined by the map
$\left([g]^{(r)}, \delta g \right) 
\rightarrow 
\left([v]^{(r)}, \eta \right)$
with
$[v]^{(r)} = \left(v^{(0)}, v^{(1)},...v^{(r-1)}\right),$
as before, and
$\eta = g^{-1}\delta {g}.$
Under the identification
$\beta_l,$
the quotient of the constraint on the variations
$C_V/G,$
becomes a subset, called {\bfseries\itshape reduced variational constraints},
$\mathfrak{C}_D \subseteq l\mathfrak{g}\oplus \mathfrak{g}.$
Since the equations
$R_K ([g]^{(k)}) = 0$
and
$R_V ([g]^{(l)}, \delta g) = 0$
are invariant, we have reduced equations
$\mathfrak{R}_K ([v]^{(k)}) = 0$
and
$\mathfrak{R}_V ([v]^{(l)}, \eta) = 0.$
Since
$R_V ([g]^{(l)}, \delta g) 
=
R_V \left([g]^{(l)}\right)\cdot\delta g$
is linear in
$\delta g,$
we have that
$\mathfrak{R}_V \left([g]^{(l)}\right)\cdot \eta$
is also linear in
$\eta.$
The Lagrangian 
$L$
gives rise to a reduced Lagrangian
$l : \mathfrak{g} \rightarrow \mathbb{R}.$
We have the following theorem
\begin{theorem}\label{redthm1}
The following conditions are equivalent
\begin{itemize}
\item[(i)]
The curve
$g(t)$
satisfies 
\[
\delta\int_{t_0}^{t_1} L(g, \dot{g})dt = 0,
\]
for all
$\delta g$
such that
$\delta g (t) \in C_V\left([g]^{(l)}(t)\right),$ for all $t \in [t_0, t_1]$ ( 
equivalently
$R_V\left([g]^{(l)}(t), \delta g (t)\right) = 0$
for all $t \in [t_0, t_1]$)
and
$\delta g (t_i) = 0$ for $i = 0, 1;$
$[g]^{(k)}(t) \in C_K$
(equivalently
$R_K\left([g]^{(k)}(t)\right) = 0$
for all $t \in [t_0, t_1]$).
\item[(ii)]
The curve
$g(t)$
satisfies the equation
\[
\left(\frac{\partial L}{\partial g}
-
\frac{d}{dt}\frac{\partial L}{\partial \dot{g}}\right)
\left([g]^{(2)}(t)\right)
\cdot \delta g
 = 0,
\]
for all
$\delta g$
such that
$\delta g (t) \in C_V\left([g]^{(l)}(t)\right),$ for all $t \in [t_0, t_1]$ 
(equivalently
$R_V\left([g]^{(l)}(t), \delta g (t)\right) = 0$
for all $t \in [t_0, t_1]$)
and
$\delta g(t_i) = 0$ for $i = 0, 1;$
$[g]^{(k)}(t) \in C_K$ 
(equivalently
$R_K\left([g]^{(k)}(t)\right) = 0$
for all $t \in [t_0, t_1]$).
\item[(iii)]
The curve
$v(t) = g^{-1}(t)\dot{g}(t)$
satisfies
\[
\delta\int_{t_0}^{t_1} l(v)dt = 0
\]
for all
$\delta v = \dot{\eta} + [v, \eta]$
where
$\eta(t) \in \mathfrak{C}_V\left([v]^{(l)}(t)\right)$ for all $t \in [t_0, t_1]$ 
(equivalently
$\mathfrak{R}_V\left([v]^{(l)}(t), \eta (t)\right) = 0$
for all $t \in [t_0, t_1]$)
and
$\eta(t_i) = 0,$ for $i = 0, 1;$
$[v]^{(k)}(t) \in \mathfrak{C}_K$
(equivalently
$\mathfrak{R}_K\left([v]^{(k)}(t)\right) = 0$
for all $t \in [t_0, t_1]$).
\item[(iv)]
The curve
$v(t) = g^{-1}(t)\dot{g}(t)$
satisfies the equation
\[
\left(-\frac{d}{dt}\frac{\partial l}{\partial v}
+
\operatorname{ad}^{\ast}\frac{\partial l}{\partial v}\right)
\left([v]^{(2)}(t)\right)
\cdot \eta
\]
for all
$\eta$
such that
$\eta(t) \in \mathfrak{C}_V\left([v]^{(l)}(t)\right)$ for all $t \in [t_0, t_1]$ 
(equivalently
$\mathfrak{R}_V\left([v]^{(l)}(t), \eta (t)\right) = 0$
for all $t \in [t_0, t_1]$)
and
$\eta(t_i) = 0,$ for $i = 0, 1;$
$[v]^{(k)}(t) \in \mathfrak{C}_K$
(equivalently
$\mathfrak{R}_K\left([v]^{(k)}(t)\right) = 0$
for all $t \in [t_0, t_1]$.)
\end{itemize}
\end{theorem}
The proof of this theorem can be performed proceeding as in \cite{CMR}. The idea 
of the proof is simple. Given a curve
$g(t)$
such that
$[g]^{(k)}(t) \in C_K$
for all $t \in [t_0, t_1]$
we take variations
$\delta g(t) = g(t)\eta (t)$
for all $t \in [t_0, t_1]$
such that
$\delta g (t) \in C_V\left([g]^{(l)}(t)\right)$ for all $t \in [t_0, t_1].$
Since
$v(t) = g^{-1}(t)\dot{g}(t)$
we can easily check that
$\delta v (t) = \eta (t) + [v(t), \eta (t)].$
The rest of the proof follows by keeping track of the reduction of both the 
kinematic constraints and the variational constraints.
\paragraph{Symmetry of the Elastic Rolling Ball.}

An interesting case occurs when, 
for each
$[g]^{(l)},$
$C_V\left([g]^{(l)}\right)$
depends only on
$g$
giving rise to a distribution
$D$
on
$G.$
This happens in the case of the rolling ball studied in 
section \ref{section2}. Let us see how the previous theorem applies to this case. 
First of all we observe that the configuration space is the
direct product group
$SO(3)\times \mathbb{R}^{2}.$
Since we are assuming an homogeneous ball the kinetic energy Lagrangian is not 
only left invariant but also right invariant. This is important because the 
constraints are also right invariant. We can thus reduce by the right action of 
the group on itself. 
For 
$\eta = (\alpha, w)$
and taking into account that the Lie bracket
in 
$\mathfrak{so}(3)$
is {\it minus} the standard one because we are reducing by 
{\it right} actions,
we have 
\begin{align}\label{redldball}
\delta \int_{t_0}^{t_1}&\left(\frac{1}{2} I \omega^2 
+ 
\frac{1}{2}M V^2\right)dt
=
0\\\label{ballred0}
\delta \omega 
&=
\dot{\alpha} - [\omega, \alpha]\\\label{ballred1}
\alpha(t_i)
&=
0,\quad\mbox{for}\quad i = 0,1\\\label{ballred2}
\delta V 
&=
\dot{w}\\\label{ballred3}
w(t_i)
&=
0,\quad\mbox{for}\quad i = 0,1\\\label{ballred4}
w 
&= 
(\alpha_2, -\alpha_1)\\\label{ballred5}
V 
&=
(\omega_2, - \omega_1)\\\label{ballred6}
\omega_3
&=
0.
\end{align}
Equations (\ref{ballred0}), (\ref{ballred2}), (\ref{ballred3}) represent the 
reduced variational constraints while equations
(\ref{ballred4}), (\ref{ballred5}), (\ref{ballred6}) represent the reduced 
kinematic constraints
(as we have explained before equation
(\ref{ballred6}) can be replaced by
$\omega_2 \dot{\omega}_1 - \omega_1 \dot{\omega}_2
=
\omega_3 (\omega_1^2 + \omega_2^2)$).
We obtain the equations of motions written in section
\ref{section2}, that is equations
(\ref{eqtnsmotion1}), (\ref{eqtnsmotion2}), (\ref{eqtnsmotion3}), (\ref{eqtnsmotion4}),
(\ref{eqtnsmotion5}).
 The reduced version of D'Alembert's Principle consists of all the previous conditions plus the condition
$\alpha_3 = 0,$
which of course corresponds to the kinematic constraint
$\omega_3 = 0.$ The D'Alembert equations are 
(\ref{eqtnsmotion1}), (\ref{eqtnsmotion2}), (\ref{eqtnsmotion4}),
(\ref{eqtnsmotion5}).
\paragraph{Rigid Ball Rolling on a Moving Plane.}
For dealing with examples where the configuration space is a principal bundle 
rather than a group and the constraints and also the Lagrangian are invariant
we need to generalize the previous theory, which we plan to do as part of future works. However, some simple examples can be worked out directly as we will see next.
Let us consider a rigid ball rolling on a plane while this plane is being 
continuously deformed according to the law
$\varphi_t : \mathbb{R}^2 \rightarrow \mathbb{R}^2.$
The Eulerian velocity is 
$v_t(x) = \dot{\varphi}_t \circ \varphi_t^{-1}(x)$
and we will assume that
$v_t(x) = v(x)$
is independent of
$t.$
For a rigid ball rolling on a fixed plane, 
that is when
$v(x) = 0,$
the system is governed by
the D'Alembert Principle which in this case is like the Principle of Virtual Work described in section \ref{section2} for an elastic ball except that one should eliminate the kinematic constraint 
$\omega_3 = 0.$
When
$v(x) \neq 0$
there is an extra force since the point 
$a$
of the ball which is in contact with the plane, is moving with velocity
$v(a),$
that is, the kinematic constraint becomes
$(\omega_2, - \omega_1) = \dot{a} - v(a).$
By differentiating with respect to
$t$
we obtain
$(\dot{\omega}_2, - \dot{\omega}_1) = \ddot{a} 
-
Dv(a).\dot{a}.$
Using this it can be easily seen that the force exerted by the floor on the ball 
is
$M\left((\dot{\omega}_2, - \dot{\omega}_1)
+
Dv(a) \cdot (\omega_2, - \omega_1) + Dv(a)\cdot v(a)
\right).$
Equations of motion can be easily derived by direct application of the basic rules of mechanics and we obtain
\begin{align}
(I + M)(\dot{\omega}_2, - \dot{\omega}_1)
&=
-MDv(a) \cdot \left[(\omega_2, - \omega_1) + v(a)\right]\\
\dot{\omega}_3
&=
0
\end{align}
Now we want to obtain the same equations using the formalism of the Principle stated in section \ref{section four}. As in the case of the elastic rolling ball this is not straightforward, which emphasizes the advantages of having a way of reducing by the symmetry as we will show next. The example under consideration is invariant with
respect to the right action of 
$SO(3)$
only because in this case the kinematic constraint is not necessarily invariant
under translations. As we have said before in this simple example a general 
theory of
reduction for systems on a principal bundle is not needed. Moreover, it is not difficult 
to prove directly that the following reduced Principle of Virtual Work gives the correct equations of motion
\begin{align}\label{redldbal7}
\delta \int_{t_0}^{t_1}
&
\left(\frac{1}{2} I \omega^2 
+ 
\frac{1}{2}M \dot{a}^2\right)dt
=
0\\\label{ballred01}
\delta \omega 
&=
\dot{\alpha} - [\omega, \alpha]\\\label{ballred11}
\alpha(t_i)
&=
0,\quad\mbox{for}\quad i = 0,1\\\label{ballred21}
\delta a
&= 
(\alpha_2, -\alpha_1)\\\label{ballred51}
(\omega_2, - \omega_1) 
&= 
\dot{a} - v(a)
\end{align}
Equations (\ref{ballred01}), (\ref{ballred11}) and 
(\ref{ballred21}) represent the variational constraints while equation
(\ref{ballred51})
is the kinematic constraint.

\

{\bf Acknowledgment}. This work has been partially supported by 
MICYT (Spain) (Grant BFM2001-2272). The work of H. Cendra was realized during a 
sabbatical year spent at Universidad Carlos III de Madrid. He also wants to thank  
CSIC 
for its kind hospitality. We all thank the referee for his helpful remarks.


\begin{thebibliography}{99}

\bibitem{Ap11}  P. Appell:
 {\it Sur les liaisons exprim\'ees par des relations
non lin\'eaires entre les vitesses}, C.R. Acad. Sci. Paris, {\bf 152} (1911), 
1197--1199;  {\it Exemple de mouvement d'un point assujetti \`a une liaison
exprim\'ee par une relation non lin\'eaire entre les composantes de la 
vittesse},
Rend. Circ. Mat. Palermo, {\bf 32} (1911), 48--50.

\bibitem{B03} A. Bloch: {\it  Non-holonomic Mechanics and Control}. 
Interdisciplinary Applied Mathematics, Springer-Verlag New York, 2003.

\bibitem{Bl96}  A. Bloch,  P.S. Krishnaprasad, J.E. Marsden, R.M. Murray: {\it
Non-holonomic Mechanical Systems with Symmetry},  Arch. Rat. Mech.
Anal. {\bf 136} (1996), 21--99.

\bibitem{Br} N.V. Brilliantov, T. P\"oschel: {\it   
  Rolling friction of a viscous sphere on a hard plane},  Europhys. Lett. {\bf 42} 
(1998), 
511--516. 


\bibitem{CLMM} F. Cantrijn, M. de Le\'on, J.C. Marrero, D. Mart{\'\i}n de Diego:
{\it Reduction of constrained systems with symmetry}, J. Math. Phys. {\bf 40} 2 
(1999), 795--820.



\bibitem{HLR}  H. Cendra, E. Lacomba and W. Reartes: {\it The Lagrange-D'Alembert-Poincare\'e Equations for the Symmetric Rolling Sphere}, VI Congreso A. Monteiro, Bahia Blanca, 2001., 19-32 Published in  2002. 


\bibitem{CMR} H. Cendra, J. E. Marsden, T.S. Ratiu: {\it Geometric mechanics, 
Lagrangian reduction and non-holonomic systems}, in Mathematics  Unlimited-2001 
and Beyond, (B. Enguist and W. Schmid, eds.), Springer-Verlag, New York (2001), 
221--273.


\bibitem{Cort} J. Cort\'es:   {\it Geometric, control and numerical aspects of 
non-holonomic systems}. Lecture 
     Notes in Mathematics, vol. 1793  Springer-Verlag 2003. 

\bibitem{Cr86} M. Crampin, W. Sarlet \& F. Cantrijn.  {\em Math. Proc. Camb.
Phil. Soc.}, {\bf 99}, 565-587 (1986).


\bibitem{Ch32} N. G. Chetaev: {\it On Gauss principle}, Izv.
Fiz-Mat. Obsc. Kazan Univ., {\bf 7} (1934), 68--71 .

\bibitem{Greidanus} J. H. Greidanus: {\it Besturing en stabiliteit van het neuswielonderstel}. Rapport V 1038. Nationaal Luchtvaartlaboratorium. Amsterdam, 1942


\bibitem{Hertz} H. Hertz: {\it \"Uber die Ber\"uhrung fester elastischer 
K\"orper}, J. reine und angew. Math. {\bf 92} (1881),  156--157. 
 


\bibitem{Ib96} A. Ibort, M. de Le\'on, G. Marmo, D. Mart\'{\i}n de Diego.  {\it
Non-holonomic constrained systems as implicit differential equations}.  Rend. Sem.
math. Univ. Pol. Torino, {\bf 54}, 3 (1996) 295-317.

\bibitem{Ko92}  J. Koiller: {\it Reduction of some classical non-holonomic
systems with symmetry}, Arch. Rational Mech. Anal., {\bf 118} (1992), 113--148.


\bibitem{Krup} O. Krupkov\'a: {\it Higher-order mechanical systems with 
constraints},  J. Math. Phys. {\bf 41} 8 (2000), 5304--5324. 

\bibitem{LL} L. D. Landau, E. M. Lifshitz, {\it Theory of Elasticity}. Butterworth-Heinemann, Oxford 1986.

\bibitem{LM1} M. de Le\'on. D. Mart{\'\i}n de Diego: {\it Solving non-holonomic
Lagrangian dynamics in terms of almost product structures},
 Extracta Mathematicae, {\bf 11} 2 (1996), 325--347.


\bibitem{LM} M. de Le\'on, D. Mart{\'\i}n de Diego:
On the geometry of non-holonomic Lagrangian systems.
{\it J. Math. Phys.} {\bf 37} (7) (1996), 3389-3414.

\bibitem{LR1} M. de Le\'on, P.R. Rodrigues:
{\it Generalized Classical Mechanics and Field Theory}.
North-Holland, Amsterdam, 1985.


\bibitem{LR2} M. de Le\'on, P.R. Rodrigues:
{\it Methods of Differential Geometry in Analytical Mechanics}.
North-Holland, Amsterdam, 1989.

\bibitem{Ma92} G. Marmo, G. Mendella, W. M. Tulczyjew:  {\it Symmetries and
constants of the motion for dynamics in implicit form}, Ann. Inst. Henri
Poincar\'e, {\bf 57} (1992), 147-166.

\bibitem{Marle}  C.-M. Marle: {\it Various approaches to conservative and 
nonconservative non-holonomic systems}, Rep. Math. Phys. {\bf 42}, 1/2 (1998) 211-
-229.


\bibitem{Me95}  G. Mendella, G. Marmo, W. Tulczyjew:  {\it
Integrability of implicit differential equations}. J. Phys. A: Math.
Gen., {\bf 28} (1995), 149--163.

\bibitem{Ne67}  J.I. Neimark, N.A. Fufaev: {\it Dynamics of Non-holonomic systems},
Translations of Mathematical Monographs, Vol. {\bf 33},  AMS Providence 1972. 


\bibitem{Pi82}  Y. Pironneau: {\it Sur les liaisons non holonomes non 
lin\'eaires
d\'eplacement virtuels \`a travail nul, conditions de Chetaev},  Proceedings of
the IUTAM--ISIMMM Symposium on ``Modern Developments in
Analytical Mechanics''.  
Eds. S. Benenti, M. Francaviglia, A. Lichnerowicz, Torino 1982, Acta
Academiae Scientiarum Taurinensis (1983), 671--686.

\bibitem{Rocard1} Y. Rocard: {\it Dynamique G{\'e}n{\'e}rale des
vibrations.} Masson et Cie. {\'E}diteurs. Paris (1949), Chap. XV, pag. 246.

\bibitem{Rocard} Y. Rocard:  {\it L'instabilit{\'e} en m{\'e}canique; automobiles, avions, ponts suspendus.} Paris, Masson, 1954.

\bibitem{DS} Do Shan: Equations of motion of systems with second-order nonlinear 
non-holonomic constraints, {\it Prikl. Mat. Mekh.} {\bf 37} (2) (1973), 349--354.


\bibitem{Va58}  V. Valcovici,  Une extension des liaisons non holonomes et
des principes variationnels,  {\it  Ber. Verh. S\"achs. Akad. Wiss. Leipzig. 
Math.-Nat. Kl.
}, {\bf 102}, 4  (1958).
 
\bibitem{Fa72}  A.M. Vershik, L.D. Faddeev:  Differential geometry and
Lagrangian mechanics with constraints, {\it Sov. Phys. Dokl.}, {\bf 17} (1972), 
34--36; 
{Lagrangian Mechanics in Invariant form}, {\it Sel. Math. Sov.}, {\bf 1} (1981), 
339--350.


 \end{thebibliography}
\end{document}